\documentclass[11pt]{article}
\usepackage{moriond,amsmath,epsfig}

\def\Journal#1#2#3#4{{#1} {\bf #2}, #3 (#4)}

\def\NPB{{\em Nucl. Phys.} B}
\def\PLB{{\em Phys. Lett.}  B}
\def\PRL{\em Phys. Rev. Lett.}
\def\PRD{{\em Phys. Rev.} D}

\def\JHEP{\em J. High Energy Phys.}
\def\EPJ{{\em Eur. Phys. J.} C}
\def\sovnp{\em Sov.~J.~Nucl.~Phys.}
\def\sovus{\em Sov.~Phys.~Usp.}
\def\yadfiz{\em Yad.~Fiz.}
\def\uspfiz{\em Usp.~Fiz.~Nauk}

\newcommand{\api}{\frac{\alpha_s}{\pi}}
\newcommand{\eqn}[1]{Eq.\,(\ref{#1})}

\def\be{\begin{equation}}
\def\ee{\end{equation}}
\def\bea{\begin{eqnarray}}
\def\eea{\end{eqnarray}}

\begin{document}
\vspace*{4cm}
\title{INCLUSIVE HIGGS BOSON PRODUCTION AT HADRON COLLIDERS AT
NEXT-TO-NEXT-TO-LEADING ORDER}

\author{\underline{W.B. KILGORE}${}^1$ and R.V. HARLANDER${}^2$}

\address{${}^{1)}$Department of Physics, Brookhaven National Lab,
Upton, New York 11973, USA\\
${}^{2)}$TH Division, CERN, CH-1211 Geneva 23, Switzerland}

\maketitle\abstracts{
Inclusive production via gluon fusion will be the most important production 
channel for Higgs boson discovery at the LHC.  I report on a calculation of 
the inclusive Higgs boson production cross section at $pp$ and $p\bar p$
colliders at next-to-next-to-leading order in QCD.
}

\section{Introduction}

The Standard Model is almost thirty-five years old, and its basic
assertion, that the weak and electromagnetic interactions unify in a
spontaneously broken $SU(2)_L\otimes U(1)_Y$ gauge theory has been
spectacularly confirmed at the quantum level by the precision
experiments at LEP and the SLC.  Still, the crucial ingredient, the
agent of electroweak symmetry breaking, remains a mystery.

The benchmark for studies of the symmetry breaking sector is the
minimal standard model, in which a single complex scalar doublet is
introduced.  Upon, spontaneous symmetry breakdown, a single Higgs
boson is left as the signature of the symmetry breaking sector.  The
final run at LEP established a 95\% confidence-level lower limit for
the Higgs boson of $114.1$ GeV~\cite{US}.  The Higgs boson can also be
constrained by its effect on precisely measured electroweak
observables.  The current best fit produced by the LEP Electroweak
Working Group predicts that $M_H = 85^{+54}_{-34}$ GeV, with a 95\%
confidence-level upper limit of $M_H < 196$ GeV.

With the end of LEP, the search for the Higgs boson is left to hadron
colliders, the Fermilab Tevatron and the CERN LHC.  Higgs production
at hadron colliders is dominated by the gluon fusion mechanism, where
gluons excite virtual top quark loops which couple strongly to the
Higgs.  Despite its dominance, this production process cannot be
exploited at the Tevatron unless the Higgs boson mass is near the $WW$
threshold.  For lighter Higgs masses, the dominant decay mode, $H\to
b\bar{b}$ is overwhelmed by the QCD background and the overall rate is
too low to permit the use of rare decay modes.  Therefore, the
Tevatron experiments must primarily rely on the associated production
mechanism, $q\bar{q}\to H+W$, for their Higgs search.

At the LHC, however, gluon fusion is the most important production
mechanism for Higgs discovery for masses below $\sim700$ GeV, since
the rate will be sufficiently high that one can use the rare decay
mode $H\to\gamma\gamma$ in the low mass region.  The discovery of the
Higgs boson and the subsequent study of its properties therefore
relies on a solid theoretical understanding of the gluon fusion
production mechanism.

Unfortunately, next-to-leading order (NLO)
studies~\cite{dawdsz,oneexact} of inclusive Higgs production do not
provide this solid understanding.  The NLO corrections are very large,
of order $70-100$\%, and initial estimates of next-to-next-to-leading
order (NNLO) corrections~\cite{KLS} based on soft plus collinear
resummation indicated that they too might be very large.  The
unsettled nature of such an important signal clearly calls for a
renewed effort to bring this process under control.

Last year, two groups presented calculations of the soft plus virtual
contributions to the NNLO correction~\cite{harkil1,CFG}.  It was known
from NLO corrections that soft plus virtual terms alone significantly
underestimate the full correction.  The leading hard scattering term
is completely collinear in nature and can be reliably obtained from
resummation.  The combined soft plus virtual plus collinear (SVC)
approximation~\cite{CFG} predicts that the total NNLO correction will
be much more moderate than the initial estimate~\cite{KLS}.  Still,
one would like a full NNLO calculation to verify that inclusive Higgs
boson production is under perturbative control.  In this talk, I will
present the results of that calculation~\cite{harkil2}.

\section{The Calculation}
In the limit that all quark masses except that of the top quark
vanish, gluons couple to Higgs only via top quark loops.  This
coupling can be approximated by an effective Lagrangian~\cite{efflag}
corresponding to the limit $m_t\to \infty$, which is valid for a large
range of $M_H$, including the currently favored region between 100 and
200\ GeV.  The effective Lagrangian is
\begin{equation}
\label{eq::efflag}
{\cal L}_{\rm eff} = -\frac{H}{4v}C_1(\alpha_s)\,G_{\mu\nu}^aG^{a\,\mu\nu}\,,
\end{equation}
where $G_{\mu\nu}^a$ is the gluon field strength tensor, $H$ is the
Higgs field, $v\approx 246$\,GeV is the vacuum expectation value of
the Higgs field and $C_1(\alpha_s)$ is the Wilson coefficient, which
for this calculation we need to order $(\alpha_s^3)$~\cite{CKS,KLS}.

\begin{figure}[h]
\hfil
(a)\psfig{figure=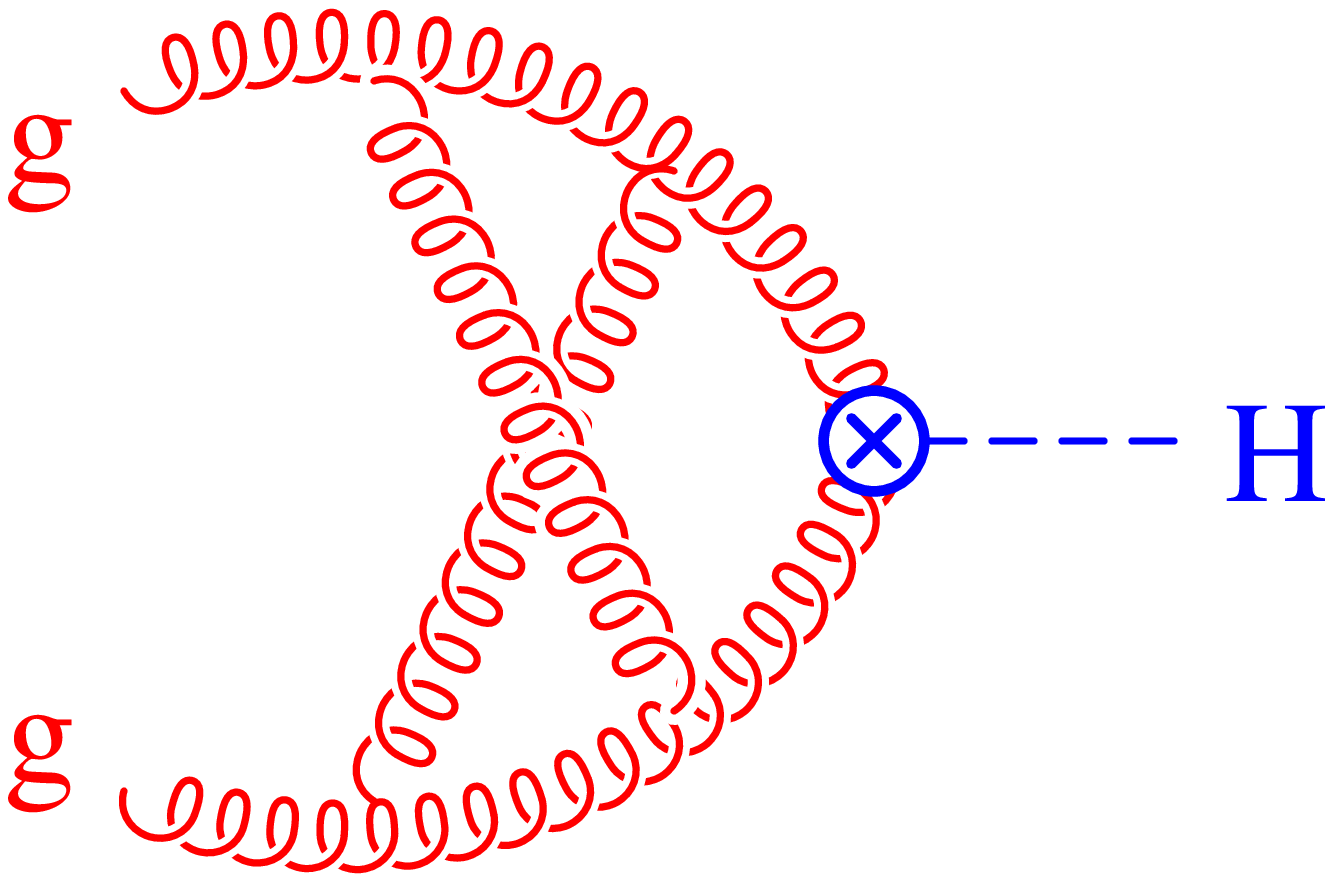,height=1in}
\hfil
(b)\psfig{figure=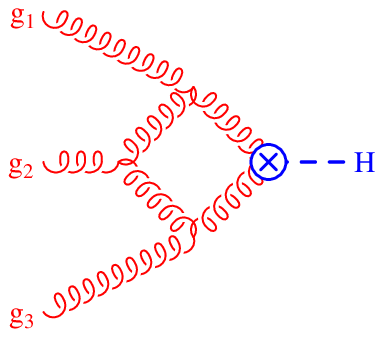,height=1in}
\hfil
(c)\psfig{figure=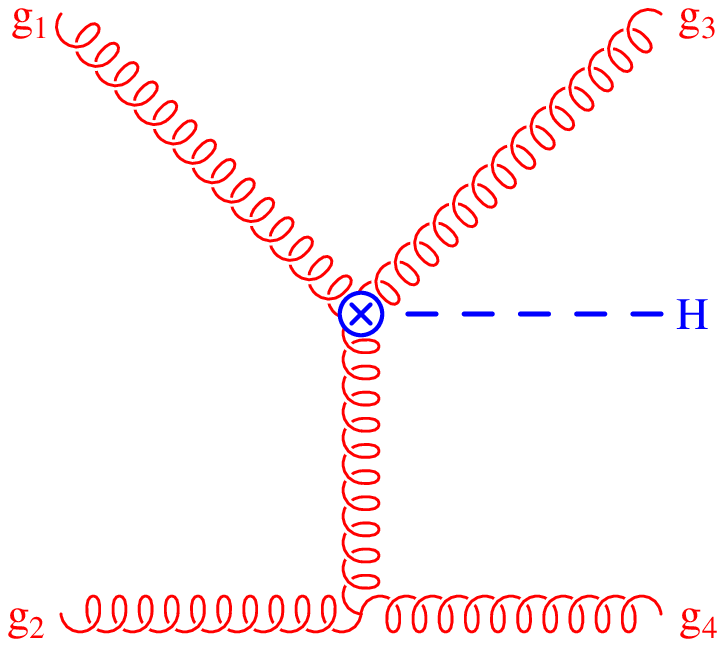,height=1in}
\hfil
(d)\psfig{figure=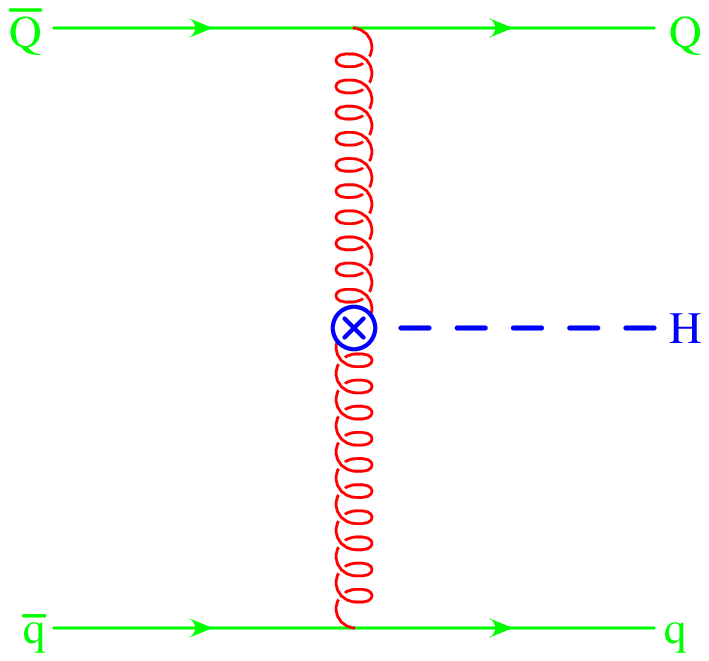,height=1in}
\hfil
\caption{\label{fig:diags}
      Sample diagrams of (a) virtual (b) single real emission and (c,d)
      double real emission corrections.}  
\end{figure}

Three classes of Feynman diagrams must be evaluated to compute the
NNLO cross section: {\rm (i)}~two-loop virtual diagrams for $gg\to H$;
{\rm (ii)}~one-loop single real emission diagrams for $gg\to Hg$,
$gq\to Hq$, and $q\bar q\to Hg$; {\rm (iii)}~tree-level double real
emission diagrams for $gg\to Hgg$, $gg\to Hq\bar q$, $gq\to Hgq$,
$qq\to Hqq$, $q\bar q\to Hgg$, and $q\bar q\to Hq\bar q$.  Sample
diagrams are shown in Figure~\ref{fig:diags}.

The most difficult part of the calculation, by far, is the computation
of the corrections due to double real emission.  While the
(tree-level) matrix elements are quite easy to compute, the
integration over double-emission phase space is quite complicated.  We
find, however, that if we expand the double real emission integrals as
power series about soft limit (where the partonic center-of-mass energy
is close to the Higgs mass, $M_H^2/\hat{s}\equiv x\to 1$), the
integrals simplify dramatically.  Moreover, the series expansion is
well-behaved and converges quite rapidly.

\begin{figure}
\begin{center}
\psfig{figure=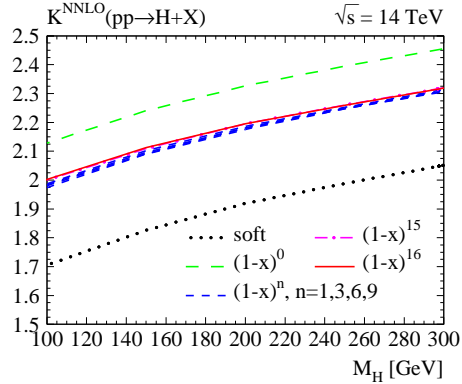,height=2in}
\caption{
      $K$-factor for Higgs production at the LHC.  Each line
      corresponds to a different order in the expansion in $(1-x)$. The
      renormalization and factorization scales are set to $M_H$.
\label{fig:nnloexp14}}
\end{center}
\end{figure}

Thus, while we have computed the virtual corrections, single real
emission and the effects of mass factorization in closed analytic
form, we have computed double real emission in the form of a power
series.  The closed-form results can also be readily expressed as
power series, so we present the full partonic cross section as an
expansion in $(1-x)$ and $\ln(1-x)$:
\begin{equation}
\begin{split}
  \hat\sigma_{ij} &= \sum_{n\geq 0}
  \left(\api\right)^n\,\hat\sigma_{ij}^{(n)}\,,\\
  \hat\sigma_{ij}^{(n)} &= a^{(n)}\,\delta(1-x) +
  \sum_{k=0}^{2n-1}b_k^{(n)}\left[\frac{\ln^k(1-x)}{1-x}\right]_+
  + \sum_{l=0}^\infty\sum_{k=0}^{2n-1}
  c_{lk}^{(n)}\, (1-x)^l\ln^k(1-x)\,,\\
  x &= \frac{M_H^2}{\hat{s}}\,.
\label{eq::abc}
\end{split}
\end{equation}
Note that if all coefficients are computed, this is an exact
expression for the partonic cross section.  In practice, we compute
only a finite number of terms.  The soft plus virtual
approximation includes only the $a^{(2)}$ and
$b^{(2)}_k$ terms at second order.  The SVC approximation
also includes the $c^{(2)}_{03}$ coefficient.  We have now computed
all coefficients $c^{(2)}_{lk}$ through $l=16$~\cite{harkil2}.  As can
be seen in Figure~\ref{fig:nnloexp14}, this is more than enough terms
to obtain a reliable result for the total cross section.

\begin{figure}[h]
\hfil
(a)\psfig{figure=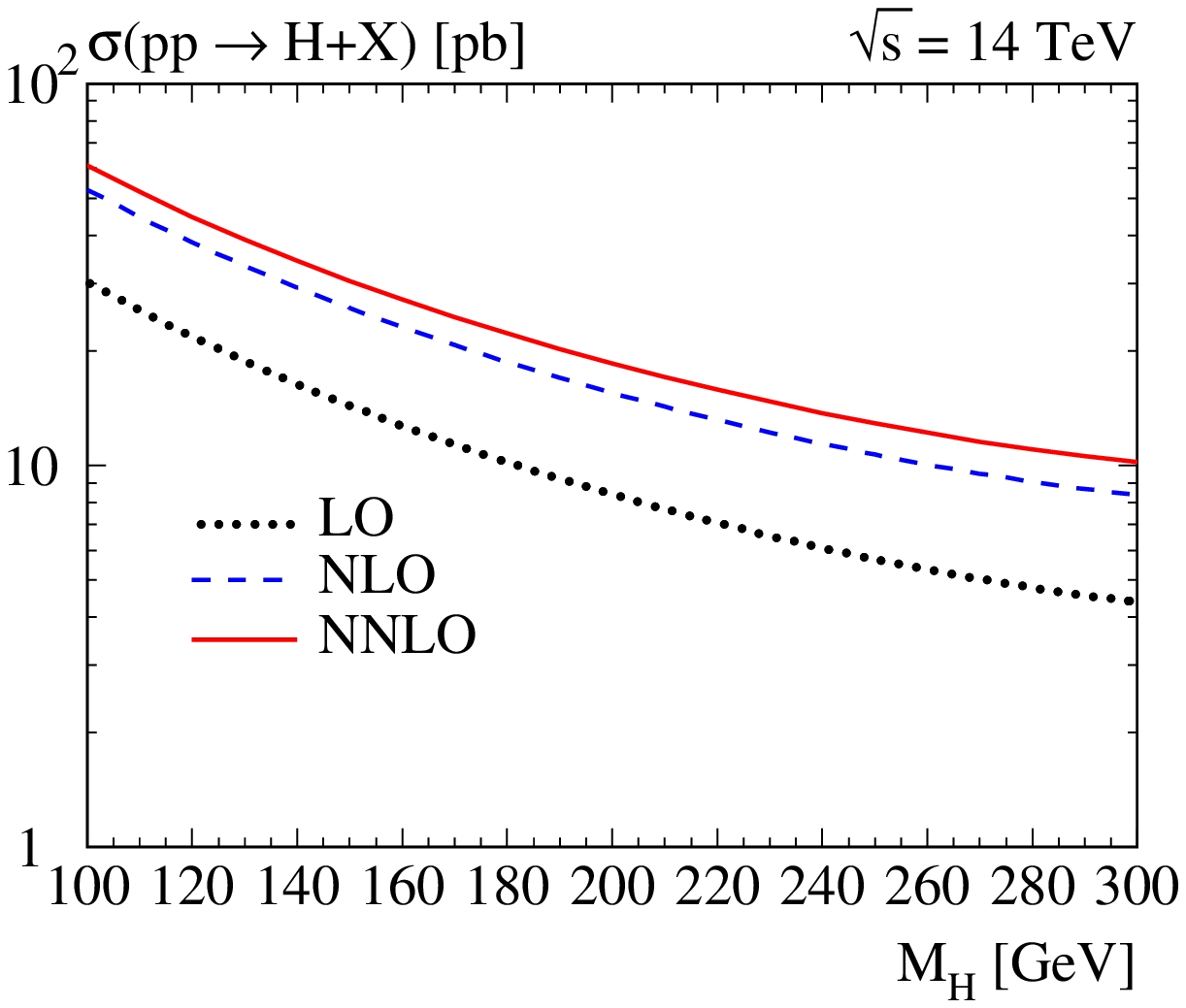,height=2in}
\hfil
(b)\psfig{figure=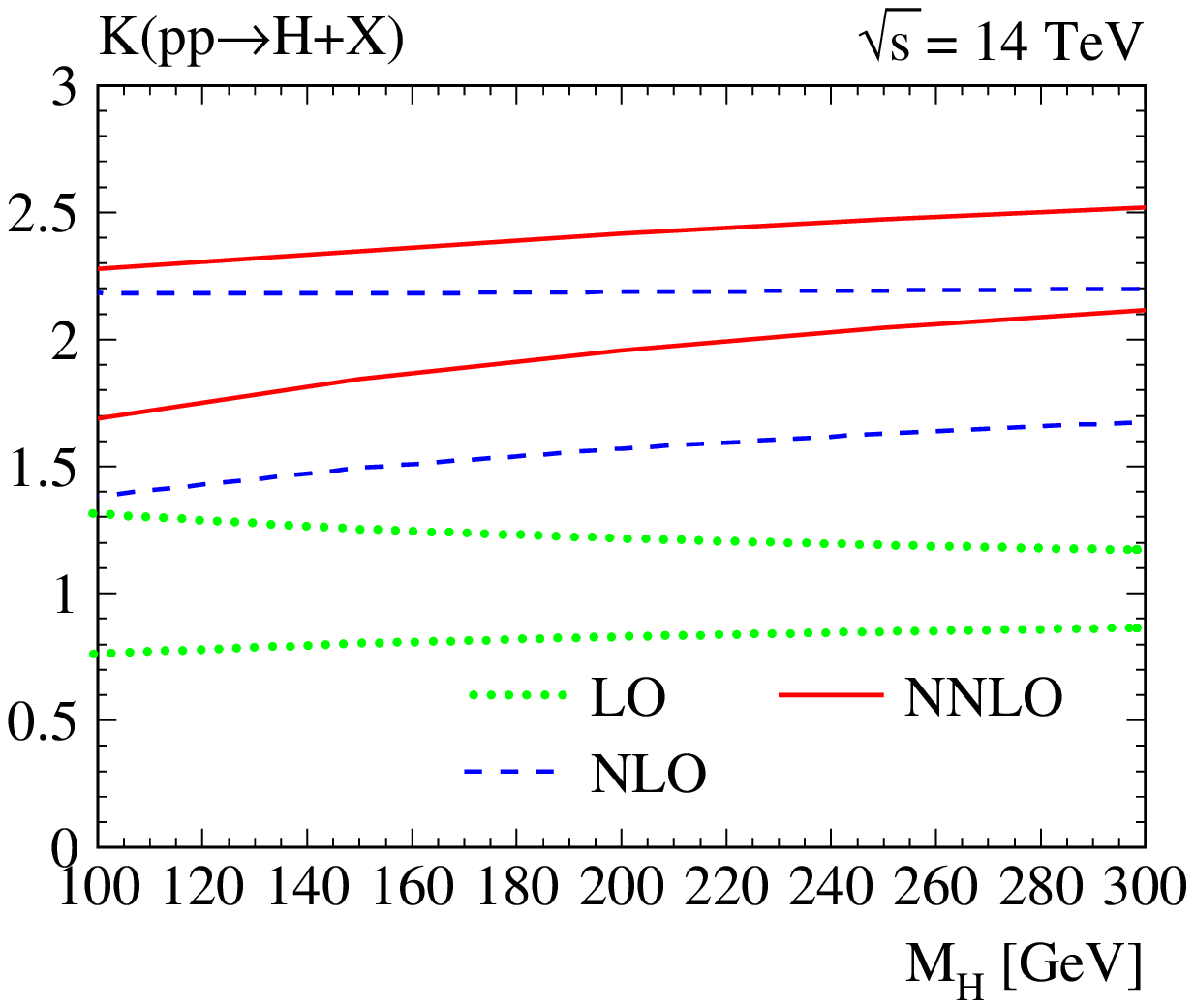,height=2in}
\hfil
\caption{
      (a) LO ({\it dotted}),
      NLO ({\it dashed}) and NNLO ({\it solid})
      cross sections for Higgs production at the LHC ($\mu_F
      = \mu_R = M_H$).  In each case, we weight the cross section
      by the ratio of the LO cross section in the full
      theory ($M_t = 175$\,GeV) to the LO cross section in
      the effective theory (\eqn{eq::efflag}).\break
      (b) Scale dependence at the LHC. The lower curve of each pair
      corresponds to $\mu_R = 2M_H$, $\mu_F=M_H/2$, the upper to
      $\mu_R =M_H/2$, $\mu_F=2M_H$.  The $K$-factor is
      computed with respect to the LO cross section at
      $\mu_R = \mu_F = M_H$.\label{fig:sigandK14}}
\end{figure}

In Figure~\ref{fig:sigandK14}(a), we show the cross section at LO, NLO
and NNLO.  At each order, we use the corresponding MRST parton
distribution set~\cite{MRST}.
One immediately sees that the true NNLO correction, while substantial,
is much smaller than the NLO correction.  Indeed, it is even a bit
smaller than predicted by the SVC approximation.  Nonetheless, it
verifies that the SVC is a good approximation of the total cross
section.

Figure~\ref{fig:sigandK14}(b) shows the renormalization and
factorization scale dependence of the ``K factor'', the ratio of the
NLO and NNLO cross sections to the leading order cross section.  The
scale dependence of the NNLO cross section is still quite large,
though somewhat smaller than at NLO.

Figure~\ref{fig:nnloexp14}\ shows the rapid convergence of the power
series expansion in $(1-x)$.  Observe that the purely soft
contributions underestimate the cross section by $\sim10\%-15\%$,
while the next term, $\propto (1-x)^0$, overestimates it by about
$5\%$.  By the time the third term in the series is included ($\propto
(1-x)^1$), one is within $1\%$ of the result obtained by computing the
first 18 terms (through $(1-x)^{16}$).  In light of the large scale
uncertainty in the result, there is little or no precision to be
gained from computing higher order terms in the expansion.

In conclusion, we have computed the full NNLO corrections to inclusive
Higgs boson production at hadron colliders.  We find reasonable
perturbative convergence and reduced scale dependence.

\section*{Acknowledgments}
The work of R.V.H. was supported in part by Deutsche
Forschungsgemeinschaft.  The work of W.B.K. was supported by the
U.~S.~Department of Energy under Contract No.~DE-AC02-98CH10886.

\end{document}